\documentclass[preprint]{elsarticle}

\usepackage[T1]{fontenc}
\usepackage[latin1]{inputenc}
\usepackage[extra]{tipa}

 \usepackage{color}

\usepackage{lineno,hyperref,amsmath,widetext}

\journal{Physics Letters B}

\begin{document}

\begin{frontmatter}

\title{Coherent gamma photon generation in a Bose-Einstein condensate of $^{135m}$Cs}

\author[ucl]{Luca Marmugi}
\author[surrey]{Philip M. Walker}
\author[ucl]{Ferruccio Renzoni\corref{mycorrespondingauthor}}
\cortext[mycorrespondingauthor]{Corresponding author}
\ead{f.renzoni@ucl.ac.uk}

\address[ucl]{Department of Physics and Astronomy, University College London, Gower Street, London WC1E 6BT, United Kingdom}
\address[surrey]{Department of Physics, University of Surrey, Guildford GU2 7XH, United Kingdom}

\begin{abstract}
We have identified a mechanism of collective nuclear de-excitation in a Bose-Einstein condensate of  $^{135}$Cs atoms in their isomeric state,  $^{135m}$Cs, suitable for the generation of coherent gamma photons. The process described here relies on coherence transfer from the Bose-Einstein condensate to the photon field, leading to collective decay triggered by spontaneous emission of a gamma photon. The mechanism differs from single-pass amplification, which cannot occur in atomic systems due to the nuclear recoil and the associated large shift between absorption and emission lines, nor does it require the large densities necessary for standard Dicke super-radiance. This overcomes the limitations that have been hindering the production of coherent gamma photons in many systems. Therefore, we propose an approach for generation of coherent gamma rays, which relies on a combination of well established techniques of nuclear and atomic physics, and can be realized with currently available technology.
\end{abstract}

\begin{keyword}
Gamma-ray lasers, Decay isomer, Bose-Einstein Condensates\\
\vskip 7pt
This article is registered under preprint number: /nucl-th/1305.0167. arXiv:1608.03468 
\end{keyword}

\end{frontmatter}


\section{Introduction}
The possibility to realize a gamma-ray laser has been an active field of research since the very first observation of lasing in the visible \cite{walker}. Possible applications range from fundamental and applied physics to the bio-medical field, the energy industry and the security sector. Numerous different mechanisms have been proposed for the generation of coherent gamma radiation, such as stimulated $\gamma$ emission from an ensemble of $^{229m}$Th  nuclei in a host  crystal \cite{tkalya}  and  annihilation of positronium  in a Bose-Einstein condensate \cite{armeniaprl, armeniapra}, to name a few.

In this work we propose coherent $\gamma$ photon generation using a Bose-Einstein condensate (BEC) of $^{135m}$Cs isomers. The use of ultra-cold atoms is attractive as it allows one to overcome two fundamental problems which have hindered the realization of a nuclear gamma-ray laser: the accumulation of a large number of isomeric nuclei, and the reduction of the gamma-ray emission linewidth, in particular of Doppler broadening, dramatically decreased at the temperature of the BEC (T$_{BEC}\sim$10$^{-7}$ K). 

However, it is not obvious which mechanisms could lead to coherent gamma ray production in such a system. Single pass amplification is inhibited by the difference in absorption and emission wavelengths, due to the large recoil associated with nuclear emission. In the specific case of the $^{135m}$Cs $\gamma$ M4 emission of interest here (see also Fig.~\ref{fig:system}), the nuclear recoil energy is $\sim$2.8 eV, while the natural linewidth of the transition, dominant at T$_{BEC}$, is $\Gamma \sim$ 10$^{-19}$~eV. This implies that, even in an ultra-cold isomeric sample, stimulated emission and amplification of spontaneously emitted photons are prevented because of Doppler shift.

Furthermore, Dicke super-radiance, a major candidate for  $\gamma$ generation \cite{baldwin1986}, cannot occur as the necessary density is not achievable in dilute atomic BECs produced via standard techniques, nor does it appear attainable in practical systems. In fact, Dicke super-radiance requires an average separation between independent emitters comparable or smaller than the wavelength $ \lambda_0$ of the radiation of interest. Hence, the  density $n_{SR}$ of the super-radiant medium must be $n_{SR} \geq \lambda_{0}^{-3}$. In the case of interest, this implies an unrealistic $^{135m}$Cs  density $n_{SR} \geq $10$^{29}$ cm$^{-3}$, corresponding to an average inter-particle separation smaller than the atomic radius. Ultimately, the characteristics of the nuclear $\gamma$ emission and the strict Dicke condition prevent the on-set of multipole-multipole correlations, which are considered the fundamental process underlying the Dicke super-radiance \cite{haroche, newpaper}.

In this Letter, we identify a mechanism for the collective de-excitation of atomic nuclei in a BEC of $^{135m}$Cs isomers leading to the generation of coherent gamma rays and we demonstrate that it can be triggered also at low atomic densities. The proposed mechanism of collective de-excitation is the atomic analogue of the collective annihilation of positronium BECs described in Refs. \cite{armeniaprl, armeniapra}. The process, which is - consistently with the above discussion - different from conventional single-pass amplification and super-radiance, takes advantage of the ultra-low temperature and the coherence of the BEC to overcome the problems of linewidth broadening, nuclear recoil and unrealistic emitters' density. Specifically, the approach proposed here relies on an absolute instability mediated by the BEC's coherence. This leads to a catastrophic decay of the BEC wavefunction, triggered by spontaneous emission from one of the trapped nuclei. The coherence of the Bose-Einstein condensate, which enables the collective nature of the de-excitation, is thus transferred to the photon field. A coherent burst of gamma photons is produced in an inverted medium constituted by an ultra-cold, coherent quantum object. In this way, no additional establishment of multipole-multipole correlations is required, in contrast with standard super-radiance occurring at high densities.

\section{Theoretical Model}
The system proposed here for gamma-ray coherent emission, unlike other approaches, has the significant advantage that it can be realized using a combination of established nuclear and atomic physics techniques. $^{135m}$Cs beams can be produced by proton-induced fission of actinides \cite{proton}. After  neutralisation \cite{neutraliser}, laser cooling and trapping, the evaporation to condensation should proceed as for the stable  $^{133}$Cs.

\begin{figure}[htbp]
	\centering
		\includegraphics[width=0.3\textwidth]{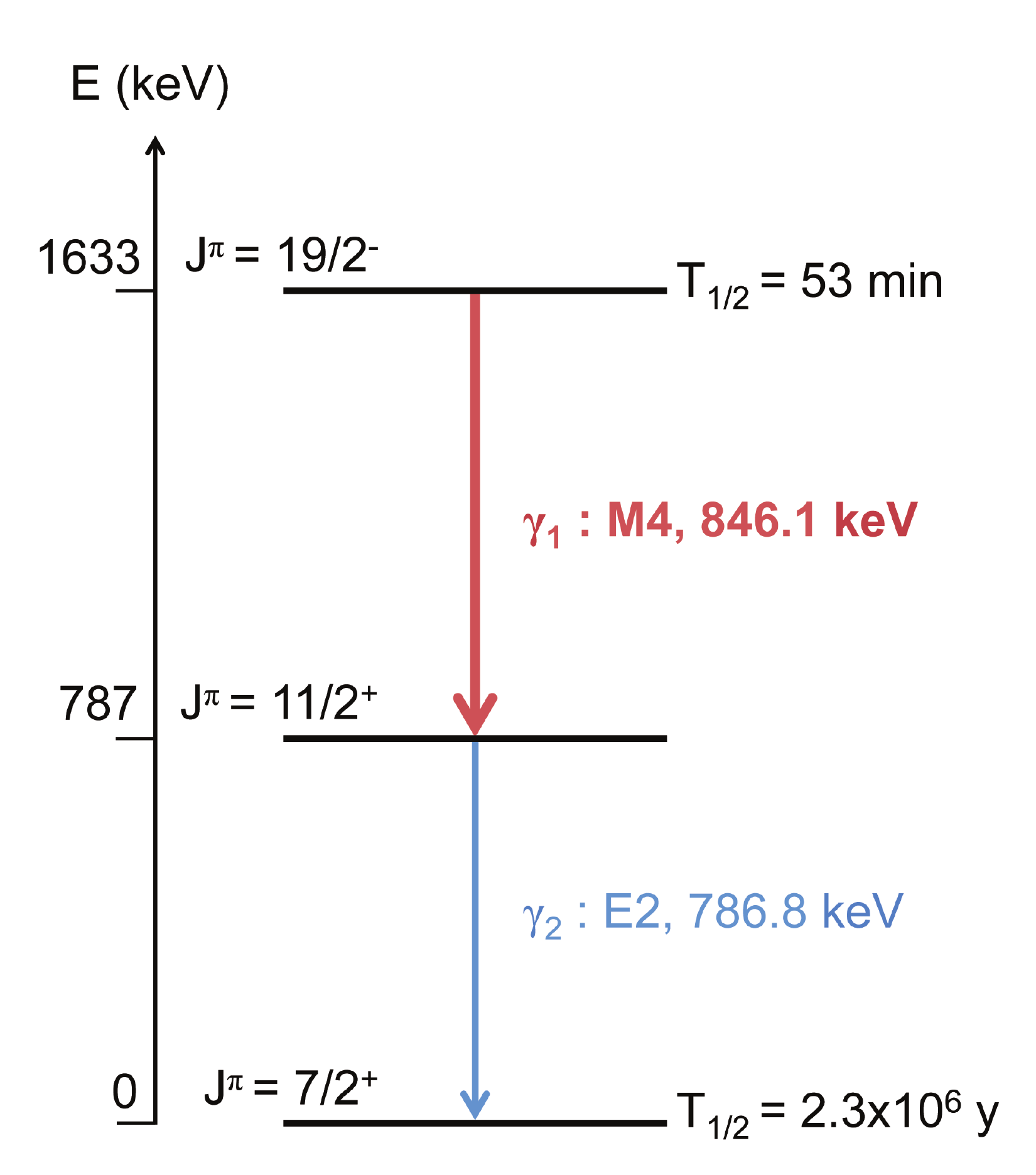}
	\caption{De-excitation scheme of  the $^{135m}$Cs  $J^{\pi}=19/2^{-}$ isomeric state. A first decay to  $11/2^{+}$ occurs via an M4 $\gamma$ transition at $846.1~keV$, followed by a  decay to the $7/2^{+}$ $^{135}$Cs ground state via an E2 transition. The $^{135}$Cs ground state has a half-life of $2.3\times 10^6$ y, thus its decay can be ignored in the context of the present study. For identifying instabilities which lead to collective decay, the analysis can be restricted to the transition $J=19/2^{-} \to J=11/2^{+}$, without taking into account the fast decay to the long-lived $J=7/2^{+}$ state.
The fast relaxation of the intermediate state $J=11/2^{+}$ can be accounted for as a broadening of the emission.
}
\label{fig:system}
\end{figure}

The nuclear system of interest is sketched in Fig.~\ref{fig:system}. It consists of an isomeric state $J^{\pi}=19/2^{-}$ decaying to $11/2^{+}$ via an M4 $\gamma$ transition at $846.1~keV$. The half-life is $T_{1/2}=53$ min. The intermediate state rapidly decays to the $7/2^{+}$ state, which corresponds to the $^{135}$Cs ground state.  It is noteworthy that the use of standard laser-cooling techniques will allow accurate trapping of only the desired $^{135m}$Cs isomer. In this way, a total population inversion, regardless the total number of trapped atoms, will be obtained.

The aim of the present work is to demonstrate the onset of instability produced by the coupling between the isomeric BEC and the photon field of the M4 transition. Therefore we first restrict our analysis to the transition $J=19/2^{-} \to J=11/2^{+}$, and  ignore subsequent fast decay to the long-lived $J=7/2^{+}$ state. Accordingly, the $^{135m}$Cs can be modelled as a two-level system, with the population inversion coinciding with the number  of isomers in the BEC.  It will then be shown that subsequent fast decay of the intermediate state $J=11/2^{+}$ to the long-lived $J=7/2^{+}$ state
leads to a broadening of the emission.
The only limitation for the mechanism at hand is the internal conversion, which represents an alternative decay channel for the  $J^{\pi}=19/2^{-}$ state, whereby the excited state dissipates energy by releasing atomic electrons. Nevertheless, the relative probability of internal conversion is $\alpha_{IC}\approx 0.04$ \cite{internalcoefficient}, thus it can be neglected for the sake of the present discussion. Therefore, within this approximation, we proceed by first assuming that the $\gamma$ decay is the only relaxation mechanism at play, with the fast relaxation of the intermediate state accounted for at a later stage. The Hamiltonian $H$ of the system can be written as the sum of three terms:

\begin{equation}
H = H_A+H_{\gamma} + H_{A\gamma}~.
\end{equation}

$H_{A}$ is the Hamiltonian of the free cesium atoms and it is defined, as in the following, in the laboratory reference frame. For the analysis of gamma-ray generation, only the nuclear excited state $J^{\pi}=19/2^{-}$ and the intermediate state  $11/2^{+}$ are considered, with the energy separation indicated  by $\hbar \omega_0=846.1~keV$.

Our theoretical model follows the standard approach to study the bulk properties of a Bose-Einstein condensate \cite{armeniaprl, armeniapra,stringari}: we consider a condensate with uniform density within a volume $V$, and then impose the limit $V \rightarrow \infty$, while keeping the BEC density constant $n_{0}$. We also  recall that, as standard in spontaneous emission processes, the wavelength of the emitted M4 radiation determines the coherence length.  Accordingly, our model applies to condensates whose size exceed all the characteristic lengths of the system.
 
By introducing second quantization operators, with the system in a cubic box of volume V with periodic boundary conditions, the atomic Hamiltonian $H_{A}$ can be written as:

\begin{equation}
H_A= \sum_{{ {\bf p}}}  \left[    \left(\frac{p^2}{2m}+{\hbar} \omega_0 \right)  \Theta^{+}_\mathbf{p}\Theta_\mathbf{p} + \left(\frac{p^2}{2m} \right)     X^{+}_\mathbf{p}X_\mathbf{p} \right] ~. 
\end{equation}

$\Theta_\mathbf{p}$ is the annihilation operator for an isomer in the excited state $J^{\pi}=19/2^{-}$, with momentum $\mathbf{p}$, and $X_\mathbf{p}$  is the annihilation  operator for the short-living intermediate state $11/2^{+}$, with momentum $\mathbf{p}$. 

The Hamiltonian of the photonic  field $H_{\gamma}$ is written as:
 
\begin{equation}
H_{\gamma} = \sum_{\mathbf{k},\zeta} {\hbar} \omega (\mathbf{k}) c^{+}_{\mathbf{k},\zeta} c_{\mathbf{k},\zeta}~,
\end{equation}

where $c_{\mathbf{k},\zeta}$ is the annihilation operator of a photon of momentum $\mathbf{k}$ and helicity $\zeta$ ($\zeta = \pm 1$), with the photonic dispersion relation $\omega=c|\mathbf{k}|$. 

Lastly, the Hamiltonian $H_{A\gamma}$

\begin{eqnarray}
H_{A\gamma} = \sum_{{\mathbf{k}},{\mathbf{p}},\zeta}  & \Big[  
{\cal M}_{\zeta}(\mathbf{k},\mathbf{p}) c^{+}_{\mathbf{k},\zeta} \Theta_\mathbf{p} X^{+}_{ {\bf p}-\hbar {\bf k}}  \nonumber \\
 & + {\cal M}_{\zeta}^{*}(\mathbf{k},\mathbf{p}) c_{\mathbf{k},\zeta} \Theta^{+}_\mathbf{p} X_{ {\bf p}- \hbar {\bf k}} \Big] 
\end{eqnarray}

describes the interaction between the isomer and the photonic field.  ${\cal M_{\zeta}}(\mathbf{k,p})$ is the amplitude of the decay of an isomer with momentum $\mathbf{p}$ into a photon of momentum $\hbar \mathbf{k}$ and an atom in the nuclear state $11/2^{+}$ with momentum $\mathbf{p}- \hbar \mathbf{k}$. In the present case, we are interested in isomers initially at rest, so we will assume $\mathbf{p=0}$ in the following.

The relevant matrix elements $|{\cal M}_{\zeta}(\mathbf{k},\mathbf{0})|^2$ can be calculated, as reported in detail in the supplemental material \cite{SImat}, from Ref.~\cite{rose}. It will suffice recalling here that  $|{\cal M}_{\zeta}(\mathbf{k},\mathbf{0})|^2$ presents explicit dependences on the initial and final states, $|J_{1,2}M_{1,2}\rangle$, and on the Euler angle $0 \leq \beta \leq \pi$ between the direction of photon emission, $\hat{\mathbf{k}}$, and the $z$-axis.

We will show now that the $^{135m}$Cs atoms, once trapped in a BEC in given conditions, exhibit absolute instability, which produces a collective decay and hence collective emission of $\gamma$ photons from phase-coupled emitters. In this sense, the key enabling factor is the coherence of the emitting medium: indistinguishable nuclei in the BEC imprint the coherence of the boson field in the photon field. In fact, this process could not happen in mere cold atomic samples. 

As initial state of the system, we assume that the photonic field is in the vacuum state, and the atoms are in a BEC of the nuclear excited state $J^{\pi}=19/2^{-}$. We  also assume that there are no atoms in the intermediate short-lived state, $11/2^{+}$, as well as in   the $7/2^{+}$ state. The latter assumption is perfectly justified, as in cold atom experiments it is possible to select which state to trap, given the isomeric shift, i.e. the difference in frequency of the D$_2$ line atomic transitions, relevant for laser cooling and trapping, between the atoms in the  nuclear excited state and in the ground state. We estimated a detuning of $\sim$0.8 GHz, roughly 10$^{2}$ times larger than the natural linewidth, between the laser cooling optical transitions of $^{135m}$Cs and $^{135}$Cs.

The system dynamics can be conveniently analyzed in the Heisenberg representation by introducing the following operators: $\tilde{c}_{\mathbf{k},\zeta} = c_{\mathbf{\mathbf{k}},\zeta} \exp[ i\omega(\mathbf{k}) t]$, $\tilde{\Theta}_\mathbf{q} = \Theta_\mathbf{q} \exp[ i (\frac{q^2}{2m \hbar}+\omega_0) t]$,  $\tilde{X}_\mathbf{q} = X_\mathbf{q} \exp[ i \frac{q^2}{2m \hbar} t]$, as described in \cite{SImat}.

We assume that the isomers' BEC constituting the atomic initial state is pure, i.e.~with condensed fraction equal to unity. In this condition, the isomers are comprised in a macroscopic quantum object of phase-coupled excited quantum oscillators. This is another striking difference with conventional super-radiance, where a classical travelling polarization oscillation sets coherent phase conditions on a number of otherwise independent emitters \cite{newpaper}. In other words, in the present case, the coherent nature of the process is an intrinsic property of the active medium and, therefore, it is extended to the whole volume occupied by the BEC.  This is formally implemented by replacing the operator $\tilde{\Theta}_\mathbf{q} $ with  the expectation number, according to the Bogoliubov c-number approximation:

\begin{equation}
\tilde{\Theta}_\mathbf{q} = \sqrt{N_0} \delta_{\mathbf{q},\mathbf{0}} ~,\label{eqn:bogoliubov}
\end{equation}

where ${N_{0}}$ is the isomeric BEC atom number and $\delta_{\mathbf{q},\mathbf{0}}$ is the Kronecker symbol.  We  do not take into account the depletion of the condensate  caused by the M4 decay. Such an  approach is suited for investigating the onset of the instability. 

By using Eq.~\ref{eqn:bogoliubov}, the system time evolution equations can be written as \cite{SImat}:

\begin{subequations}
\begin{eqnarray}
 i \hbar \dot{\tilde{c}}_{\mathbf{k},\zeta}           &=&  \sqrt{n_0}  \xi_{\zeta} (\mathbf{k},\mathbf{0}) \tilde{X}^{+}_\mathbf{-\hbar k} \exp[ i \Delta_\mathbf{ k} t ] \label{eqn:ctildedot} \\
 i\hbar \dot{\tilde{X}}_{\hbar \mathbf{k}}       &=&  \sqrt{n_0}  \sum_{\zeta}  \xi_\zeta (\mathbf{-k},\mathbf{0})  \tilde{c}^{+}_{\mathbf{-k},\zeta}    \exp[i \Delta_\mathbf{- k} t ]~ .
\end{eqnarray} \label{eq:red_eq}
\end{subequations}

Here $n_0=N_0/V$ is the volume density of the isomeric BEC,  $\xi_{\pm} (\mathbf{k},0)\equiv\sqrt{V}{\cal M}_{\pm}(\mathbf{k},0)$, which makes the results independent of the BEC volume, and:

\begin{equation}
\Delta_{\pm \mathbf{k}} =  \omega(\mathbf{\pm  k})+\dfrac{\hbar (\pm k)^2}{2m}-\omega_0~.\label{eqn:delta}
\end{equation}

In Eq.~\ref{eqn:delta} the opposite linear momenta $\hbar \mathbf{ k}$ and $-\hbar \mathbf{ k}$ of the photon and the ground state of the M4 transition are made explicit.
 
The remaining explicit time-dependence in Eqs. \ref{eq:red_eq} can be eliminated by introducing the operator
\begin{equation}
\doubletilde{c}_{\mathbf{k},\zeta} = \tilde{c}_{\mathbf{k},\zeta} \exp\left[- i\Delta_{ {\mathbf k}}      t\right]~,
\end{equation}
 so that the relevant equations become
\begin{subequations}
\begin{eqnarray}
 i \hbar \dot{\doubletilde{c}}_{\mathbf{k},\zeta}           &=&  \hbar\Delta_\mathbf{k}\doubletilde{c}_{\mathbf{k},\zeta}+\sqrt{n_0}  \xi_{\zeta} (\mathbf{k},\mathbf{0}) \tilde{X}^{+}_{{ -\hbar {\bf k}}}  \label{eqn:ctildedot} \\
 i\hbar \dot{\tilde{X}}_{\hbar \mathbf{k}}       &=&  \sqrt{n_0}  \sum_{\zeta}  \xi_\zeta (\mathbf{-k},\mathbf{0})  \doubletilde{c}^{+}_{\mathbf{-k},\zeta}~.     
\end{eqnarray} \label{eq:red_eq_bis}
\end{subequations}

\subsection{Critical Instabilty}
From Eqs.~\ref{eq:red_eq_bis}, we calculate \cite{SImat}  the total number of  photons  $N_{\bf k }(t)$ emitted in the ${\bf k}$-mode:
\begin{equation}
N_{\bf k} (t) = N_{{\bf k} ,-}+N_{{\bf k},+} =  \frac{2n_0( |\xi_{-} (\mathbf{k},\mathbf{0})|^2 + |\xi_{+} (\mathbf{k},\mathbf{0})|^2)    \left[-\cos(\delta_0 t)+\cosh(\delta_1 t)\right]}{\hbar^2 (\delta_0^2+\delta_1^2)},~
\end{equation}

where the index $\pm$ refers to the photon helicity, and

\begin{equation}
\delta_0+i \delta_1 = \frac{1}{\hbar} \sqrt{ \hbar^2\Delta_{\mathbf{k}}^2 - 4 n_0  ( |\xi_{-} (\mathbf{k},\mathbf{0})|^2 + |\xi_{+} (\mathbf{k},\mathbf{0})|^2) }~.
\end{equation}

 This produces an exponential emission rate for sufficiently large $n_0$. The condition for exponential  growth of $N_{\bf k}$ is given by $\delta_{1}\neq0$, which requires:

\begin{equation}
 \hbar^2\Delta_{\mathbf{k}}^2 -  4 n_0 ( |\xi_{-} (\mathbf{k},0)|^2 + |\xi_{+} (\mathbf{k},\mathbf{0})|^2)< 0~.
\end{equation}

\begin{figure}[htbp]
	\centering
	\includegraphics[width=0.45\textwidth]{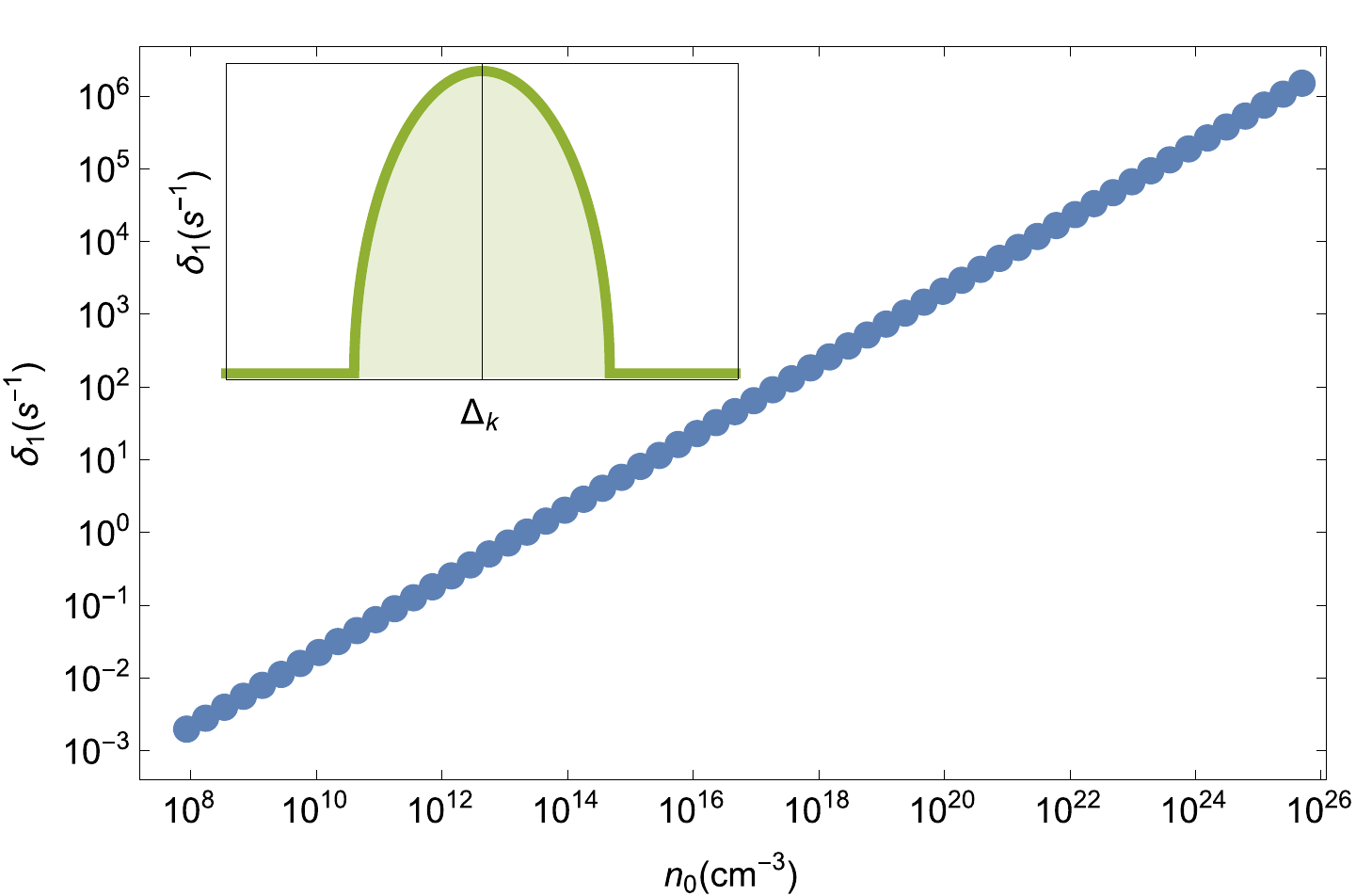}
	\caption{
Critical parameter, averaged over the initial $M_{1}$ states and summed over the final ones, as a function of the atomic density $n_0$. Inset: Absolute instability region, with average  $\delta_{1}$ plotted as a function of $\Delta_{\bf k}$, evidencing  collective emission in the interval defined by Eq.~(\ref{gain_interval}).}
	\label{fig:gain}
\end{figure}

Thus,  exponential photonic generation is observed in the interval of frequencies (see also inset of Fig.~\ref{fig:gain}):

\begin{equation}
 -\frac{2\sqrt{n_0 }}{\hbar}\sqrt{\sum_{\zeta}|\xi_\zeta(\mathbf{k},\mathbf{0})|^2}  < \Delta_{\bf k} <   \frac{2\sqrt{n_0 }}{\hbar}\sqrt{\sum_{\zeta}|\xi_\zeta(\mathbf{k},\mathbf{0})|^2} ~. \label{gain_interval}
\end{equation}

At the center emission frequency $\Delta_{\mathbf {k}} = 0$, i.e. for emission frequency $\omega = \omega_0-\omega_R$ with $\omega_R= \hbar k^2/(2m)$ the recoil frequency, the critical parameter $\delta_1$ is:

\begin{equation}
\delta_1 {\Big|_{\Delta_{\mathbf {k}} = 0}}=  \frac{2\sqrt{n_0}}{\hbar} \sqrt{\sum_{\zeta}|\xi_\zeta(\mathbf{k},\mathbf{0})|^2}~.
\end{equation}

$\delta_{1}$  exhibits a $\sqrt{n_0}$ dependence, and it is non-zero also at low atomic densities, such as those currently attainable in an atomic BEC.

The critical parameter depends on the initial and final $M$-states, as well as on the emission angle $\beta$. We thus consider $\delta_{1}$ averaged over the initial $M$-states and summed over the final ones, which becomes independent of $\beta$. Its dependence on the  isomeric BEC atomic density $n_0$ is reported in Fig.~\ref{fig:gain}, as well as a plot of the absolute instability region, where collective emission of coherent $\gamma$ photons happens.

The behavior exhibited in Fig.~\ref{fig:gain} - in the gamma spectral region - is possible only because of the coherence of the BEC, which eliminates the need of initial spontaneous oscillations of the fields propagating through the medium,  characterized by specific delay. In other words, the fact that all the excited quantum oscillators are already comprised of a single wavefunction (Eq.~\ref{eqn:bogoliubov}) automatically establishes stable phase-coupling among all the emitters and thus provides the conditions to overcome the Dicke limit.

As consequence, the process is dramatically different from single-pass amplification, where a ``seed'' quantum of the photonic field is coherently amplified by the surrounding medium. Here, the spontaneous decay of one isomer in the BEC - provided that condition Eq.~\ref{gain_interval} is satisfied -  triggers an absolute instability, independent of  the dissipation regime and the shape of the emitting particles. This results in a collective decay of the isomers and in the consequent emission of a coherent pulse of 846.1 keV photons and collapse of the BEC. 

The scattering length of $^{135m}$Cs and the details of the collisional processes at ultra-cold temperatures are not currently known, therefore a precise estimate of the details of the condensation process or of the expected final atom number $ N_{0}$ is - to date - not possible. Nevertheless, with a condensed fraction equal to unity, one could infer from results from stable cesium that, with typical densities and number of trapped isomers in the BEC, a burst of 10$^{4}$-10$^{5}$ coherent photons will be obtained, as produced by a BEC of similar atom number. In actual experiments, after production, electrostatic acceleration, mass separation and neutralization of the desired $^{135m}$Cs, atoms can be trapped in a magneto-optical trap (MOT), via standard laser cooling techniques. The MOT will allow also a further purification of the sample and a preliminary reduction of the linewidth, thanks to the suppression of Doppler broadening. At this stage, however, the lack of coherence prevents any possibility to observe collective phenomena. The atomic cloud will be then transferred to a far-detuned optical dipole trap, where forced evaporation will lead to quantum degeneracy and, ultimately, to a pure BEC. Here, at around 10$^{-7}$ K and 10$^{12}$ cm$^{-3} \leq n_{0} \leq$ 10$^{14}$ cm$^{-3}$, the conditions for coherent generation of $\gamma$ photons will be satisfied within the instability region (Eq.~\ref{gain_interval}). At this stage, absorption imaging of the atomic cloud performed on the optical D$_{2}$ transition with conventional infra-red laser diodes and the investigation of sample displacement produced by the collective recoil will provide unambiguous indication of the onset of the collective decay. Although internal conversion is expected to reduce the number of significant events, in this case the cloud recoil would be negligible with respect to that produced by the coherent $\gamma$ emission.

\begin{figure}[htbp]
	\centering
	\includegraphics[width=0.45\textwidth]{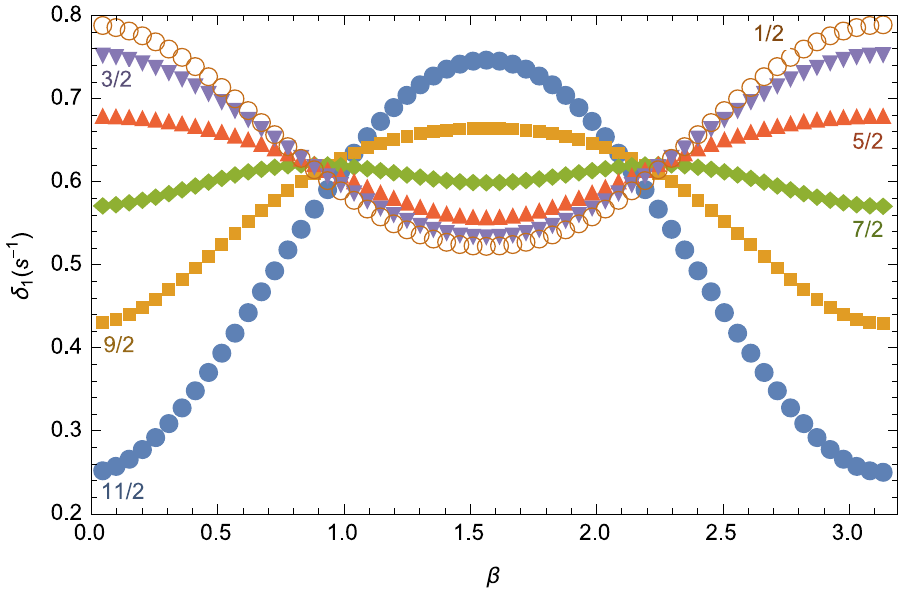}
	\caption{
Critical parameter, averaged over the initial $M$ states, as a function of the emission angle $\beta$, for different final $M$-states and for  an isomeric BEC atomic density $n_0 = 10^{14}$  cm$^{-3}$.
}
	\label{fig:average}
\end{figure}

Moreover, the gamma-ray emission exhibits a non-trivial angular distribution, as a consequence of the dependence of $|\xi_{\zeta}(\mathbf{k}, \mathbf{0})|^{2}$ on the final $M_{2}$ states.  To investigate such a distribution, which could be in principle used to further investigate the coherent $\gamma$ emission, in Fig.~\ref{fig:average} the dependence of the critical parameter on $\beta$ is displayed for an average over the initial state $M_1$ and a specific choice of the final state $M_2$.

So far we restricted our analysis to a closed two-level system consisting of the nuclear excited state $J^{\pi}=19/2^{-}$ and  the state  $11/2^{+}$.  Our model can be generalised to include the fast relaxation of the  intermediate state \cite{SImat}. The collective mechanism  identified here still holds in the presence of the relaxation of the intermediate state, with the latter one resulting into a broadening of the emission over a range a frequencies determined by the intermediate state relaxation rate. 

\section{Conclusions}
Our results demonstrate that a mechanism of collective decay occurs for a  BEC of  $^{135m}$Cs isomers. The collective nature of the phenomenon is highlighted by the exponential dependence of the number of emitted photons with respect to the initial isomer density. The collective de-excitation relies on the coherence of the condensate being transferred to $\gamma$ photons, and occurs at densities much lower than those required by the standard Dicke super-radiance. The identified mechanism provides a promising route to the generation of coherent $\gamma$ radiation, as the associated process can be realized with available technology.  $^{135m}$Cs ion beams can be generated by proton-induced fission of actinides. Afterwards, laser cooling and trapping can proceed as well established for $^{133}$Cs and some of its isotopes \cite{motiso}. The long lifetime of $^{135m}$Cs allows for evaporation and creation of a BEC in an optical trap, along the lines of the procedure for stable cesium. As the collisional properties of ultra-cold $^{135m}$Cs are not known, it is not possible to give an accurate estimate of the expected size of the BEC, and hence of the intensity of the $\gamma$ photons burst. However, the present results indicate that exponential photonic generation occurs for a wide range of BEC densities. We therefore expect coherent emission to occur over a broad range of BEC size, thus demonstrating the validity of the proposed approach for coherent gamma-ray generation.

\section*{Acknowledegments}
This work was partially funded by the H2020-EU$.1.3.2$  programme through the Marie Curie Fellowship 2020-MSCA-IF-2014 ``GAMMALAS'' to L.~M.  (Proj. Ref. 657188), and by the Royal Society. P.~M.~W. acknowledges STFC support under grant no. ST/L005743/1.

\section*{References}

\bibliography{isomers}

\begin{thebibliography}{10}
\expandafter\ifx\csname url\endcsname\relax
  \def\url#1{\texttt{#1}}\fi
\expandafter\ifx\csname urlprefix\endcsname\relax\def\urlprefix{URL }\fi
\expandafter\ifx\csname href\endcsname\relax
  \def\href#1#2{#2} \def\path#1{#1}\fi

\bibitem{walker}
P.~Walker, G.~Dracoulis, \href{http://dx.doi.org/10.1038/19911}{Energy traps in
  atomic nuclei}, Nature 399~(6731) (1999) 35--40.
\newline\urlprefix\url{http://dx.doi.org/10.1038/19911}

\bibitem{tkalya}
E.~V. Tkalya,
  \href{http://link.aps.org/doi/10.1103/PhysRevLett.106.162501}{Proposal for a
  nuclear gamma-ray laser of optical range}, Phys. Rev. Lett. 106 (2011)
  162501.
\newblock \href {http://dx.doi.org/10.1103/PhysRevLett.106.162501}
  {\path{doi:10.1103/PhysRevLett.106.162501}}.
\newline\urlprefix\url{http://link.aps.org/doi/10.1103/PhysRevLett.106.162501}

\bibitem{armeniaprl}
H.~K. Avetissian, A.~K. Avetissian, G.~F. Mkrtchian,
  \href{http://link.aps.org/doi/10.1103/PhysRevLett.113.023904}{Self-amplified
  gamma-ray laser on positronium atoms from a {B}ose-{E}instein condensate},
  Phys. Rev. Lett. 113 (2014) 023904.
\newblock \href {http://dx.doi.org/10.1103/PhysRevLett.113.023904}
  {\path{doi:10.1103/PhysRevLett.113.023904}}.
\newline\urlprefix\url{http://link.aps.org/doi/10.1103/PhysRevLett.113.023904}

\bibitem{armeniapra}
H.~K. Avetissian, A.~K. Avetissian, G.~F. Mkrtchian,
  \href{http://link.aps.org/doi/10.1103/PhysRevA.92.023820}{Gamma-ray laser
  based on the collective decay of positronium atoms in a {B}ose-{E}instein
  condensate}, Phys. Rev. A 92 (2015) 023820.
\newblock \href {http://dx.doi.org/10.1103/PhysRevA.92.023820}
  {\path{doi:10.1103/PhysRevA.92.023820}}.
\newline\urlprefix\url{http://link.aps.org/doi/10.1103/PhysRevA.92.023820}

\bibitem{baldwin1986}
G.~C. Baldwin, M.~S. Feld,
  \href{http://scitation.aip.org/content/aip/journal/jap/59/11/10.1063/1.336747}{Kinetics
  of nuclear super-radiance}, Journal of Applied Physics 59~(11) (1986)
  3665--3671.
\newblock \href {http://dx.doi.org/http://dx.doi.org/10.1063/1.336747}
  {\path{doi:http://dx.doi.org/10.1063/1.336747}}.
\newline\urlprefix\url{http://scitation.aip.org/content/aip/journal/jap/59/11/10.1063/1.336747}

\bibitem{haroche}
M.~Gross, S.~Haroche,
  \href{http://www.sciencedirect.com/science/article/pii/0370157382901028}{Superradiance:
  An essay on the theory of collective spontaneous emission}, Physics Reports
  93~(5) (1982) 301 -- 396.
\newblock \href
  {http://dx.doi.org/http://dx.doi.org/10.1016/0370-1573(82)90102-8}
  {\path{doi:http://dx.doi.org/10.1016/0370-1573(82)90102-8}}.
\newline\urlprefix\url{http://www.sciencedirect.com/science/article/pii/0370157382901028}

\bibitem{newpaper}
V.~V. Zheleznyakov, V.~V. Kocharovskiĭ, V.~V. Kocharovskiĭ,
  \href{http://stacks.iop.org/0038-5670/32/i=10/a=R01}{Polarization waves and
  super-radiance in active media}, Soviet Physics Uspekhi 32~(10) (1989) 835.
\newline\urlprefix\url{http://stacks.iop.org/0038-5670/32/i=10/a=R01}

\bibitem{proton}
B.~Tracy, J.~Chaumont, R.~Klapisch, A.~Nitschke, J.M.~Poskanzer, E.~Roeckl,
  C.~Thibault, {R}b and {C}s isotopic cross sections from 40-60-{M}ev-proton
  fission of $^{238}${U}, $^{232}${Th}, and $^{235}${U}, Phys. Rev. C 5 (1972)
  222--234.

\bibitem{neutraliser}
S.~Aubin, E.~Gomez, L.~Orozco, G.~Sprouse, High efficiency magneto-optical trap
  for unstable isotopes, Rev. Sci. Instr. 74 (2003) 4342--4351.

\bibitem{internalcoefficient}
T.~Kib\'edi, T.~Burrows, M.~Trzhaskovskaya, P.~Davidson, C.~N. Jr.,
  \href{http://www.sciencedirect.com/science/article/pii/S0168900208002520}{Evaluation
  of theoretical conversion coefficients using {B}r{I}cc}, Nuclear Instruments
  and Methods in Physics Research Section A: Accelerators, Spectrometers,
  Detectors and Associated Equipment 589~(2) (2008) 202 -- 229.
\newblock \href
  {http://dx.doi.org/http://dx.doi.org/10.1016/j.nima.2008.02.051}
  {\path{doi:http://dx.doi.org/10.1016/j.nima.2008.02.051}}.
\newline\urlprefix\url{http://www.sciencedirect.com/science/article/pii/S0168900208002520}

\bibitem{stringari}
L.~Pitaevskii, S.~Stringari, Bose-einstein condensation, Oxford Science
  Publications.

\bibitem{SImat}
See supplemental material at [url].
\newblock \href{http://www.url.ext/}{[link]}.
\newline\urlprefix\url{http://www.url.ext/}

\bibitem{rose}
H.~J. Rose, D.~M. Brink,
  \href{http://link.aps.org/doi/10.1103/RevModPhys.39.306}{Angular
  distributions of gamma rays in terms of phase-defined reduced matrix
  elements}, Rev. Mod. Phys. 39 (1967) 306--347.
\newblock \href {http://dx.doi.org/10.1103/RevModPhys.39.306}
  {\path{doi:10.1103/RevModPhys.39.306}}.
\newline\urlprefix\url{http://link.aps.org/doi/10.1103/RevModPhys.39.306}

\bibitem{motiso}
M.~Di~Rosa, S.~Crane, J.~Kitten, W.~Taylor, D.~Vieira, X.~Zhao, Magneto-optical
  trap and mass-separator system for the ultra-sensitive detection of
  $^{135}${C}s and $^{137}${C}s, Appl. Phys. B 76 (2003) 45--55.

\end{thebibliography}


\begin{thebibliography}{}
\expandafter\ifx\csname url\endcsname\relax
  \def\url#1{\texttt{#1}}\fi
\expandafter\ifx\csname urlprefix\endcsname\relax\def\urlprefix{URL }\fi
\expandafter\ifx\csname href\endcsname\relax
  \def\href#1#2{#2} \def\path#1{#1}\fi

\end{thebibliography}

\end{document}


\begin{frontmatter}

\title{Supplemental material for\\ ``Coherent gamma photon generation in a Bose-Einstein condensate of $^{135m}$Cs''}

\author[ucl]{Luca Marmugi}
\author[surrey]{Philip M. Walker}
\author[ucl]{Ferruccio Renzoni\corref{mycorrespondingauthor}}
\cortext[mycorrespondingauthor]{Corresponding author}
\ead{f.renzoni@ucl.ac.uk}

\address[ucl]{Department of Physics and Astronomy, University College London, Gower Street, London WC1E 6BT, United Kingdom}
\address[surrey]{Department of Physics, University of Surrey, Guildford GU2 7XH, United Kingdom}
\end{frontmatter}

\section{Matrix element manipulations}
In this section, we derive the matrix element for the M4 operator, used in the main text \cite{main}. 

The starting point is the expression for the matrix element of  the magnetic operator  M4 \cite{m4}:

\begin{equation}
{\cal M}_\zeta (\mathbf{k},\mathbf{0}) = - \sqrt{\dfrac{2\pi\hbar c}{Vk}} \zeta  \sum_M \langle  J_1 M_1 | T_{LM}| J_2 M_2  \rangle {\cal D}_{M\zeta}^{L*}(R) \label{eqn:em}
\end{equation}

with $L=4$, and where  ${\cal D}^L_{M\zeta}(R)$ is the Wigner D-matrix for a rotation $R=(\alpha,\beta,\gamma)$ which takes the $z$-axis to the direction of $\mathbf{k}$. It is noteworthy that only $|{\cal M}_\zeta (\mathbf{k},\mathbf{0})|^2$ enters the relevant expressions in the analysis presented in the main text \cite{main}.

Using the Wigner theorem, it is possible to express the matrix element of the operator $T_{LM}$, connecting the two states $|J_{1},M_{1}\rangle$ and $|J_{2},M_{2}\rangle$, via Clebsch-Gordan (CG) coefficients:

\begin{equation}
\langle  J_1 M_1 | T_{LM}| J_2 M_2  \rangle  = (-1)^{2L} \langle J_2 M_2 L M|J_1M_1\rangle \langle  J_1 || T_L|| J_2  \rangle ~.
\end{equation}

The reduced matrix element $ \langle  J_1 || T_L|| J_2  \rangle$ can be calculated from the known mean life $\tau=T_{1/2}/\ln 2$ of the excited state:

\begin{equation}
\frac{1}{\tau}=\frac{4k}{\hbar} \frac{|\langle  J_1 || T_L|| J_2  \rangle|^2}{2L+1}~. \label{eqn:lifetime}
\end{equation}

Thus, by using $(-1)^{4L|_{L=4}}=+1$ and taking into consideration that, for the two helicity states $\zeta^{2}=(\pm1)^{2}=1$, Eq.~\ref{eqn:em} can be written as:

\begin{equation}
|{\cal M}_{\zeta}(\mathbf{k},\mathbf{0})|^2 = \frac{2\pi\hbar c}{Vk}|\langle J_1||T_L||J_2\rangle|^2  \sum_{MM^\prime}  \langle J_2 M_2 L M|J_1M_1 \rangle \langle J_2 M_2 L M^\prime|J_1M_1 \rangle
{\cal D}^{L*}_{M\zeta} {\cal D}^{L}_{M^\prime \zeta ~.}
\end{equation}

Now, by using the property of the Wigner matrices:

\begin{equation}
{\cal D}^{j*}_{m,n}(\alpha,\beta,\gamma) = (-1)^{m-n} {\cal D}^{j}_{-m,-n}(\alpha,\beta,\gamma) \label{eqn:wignerproperty}~,
\end{equation}

we have:

\begin{equation}
|{\cal M}_{\zeta}(\mathbf{k},\mathbf{0})|^2 = \frac{2\pi\hbar c}{Vk}|\langle J_1||T_L||J_2\rangle|^2   \sum_{MM^\prime} \langle J_2 M_2 L M|J_1M_1 \rangle \langle J_2 M_2 L M^\prime|J_1M_1 \rangle
(-1)^{M-\zeta}{\cal D}^{L}_{-M-\zeta} {\cal D}^{L}_{M^\prime \zeta}~.
\end{equation}

The Wigner matrices in the above equation can be further simplified by referring to the following general property:

\begin{equation}
{\cal D}^{j_1}_{m_1,n_1}(\alpha,\beta,\gamma){\cal D}^{j_2}_{m_2,n_2}(\alpha,\beta,\gamma) = \sum_{K=|j_1-j_2|}^{j_1+j_2}\sum_{H=-K}^K\sum_{N=-K}^K \langle j_1 m_1j_2 m_2|KH\rangle \langle j_1n_1j_2 n_2|KN \rangle {\cal D}^{K}_{H,N}(\alpha,\beta,\gamma)~. \label{eqn:generalproperty}
\end{equation}

Thanks to Eq.~\ref{eqn:generalproperty}, we thus obtain:

\begin{eqnarray}
|{\cal M}_{\zeta}(\mathbf{k},\mathbf{0})|^2 &=& \frac{2\pi\hbar c}{Vk}|\langle J_1||T_L||J_2\rangle|^2\sum_{MM^\prime} \langle J_2 M_2 L M|J_1M_1\rangle\langle J_2 M_2 L M^\prime|J_1M_1\rangle (-1)^{M-\zeta}\times \nonumber \\
&&\sum_{K=0}^{2L}\sum_{H=-K}^K \sum_{N=-K}^K \langle L-MLM^\prime|KH\rangle \langle L-\zeta L \zeta|KN\rangle  {\cal D}^{K}_{HN}~. \label{eqn:general}
\end{eqnarray}

Here, the only non-zero CG coefficients are  for $M^\prime=M$, $N=0$ and $H=0$. Thus, the previous expression becomes:

\begin{equation}
|{\cal M}_{\zeta}(\mathbf{k},\mathbf{0})|^2 = \frac{2\pi\hbar c}{Vk}|\langle J_1||T_L||J_2\rangle|^2  \sum_{M} \langle J_2 M_2 L M|J_1M_1\rangle^2  (-1)^{M-\zeta} \sum_{K=0}^{2L} \langle L-MLM|K0\rangle \langle L-\zeta L \zeta|K0\rangle {\cal D}^{K}_{00}~. \label{eqn:intermediate}
\end{equation}

We note that, with respect to the sum over $M$, the only non zero CG in Eq.~\ref{eqn:intermediate} is the one for $M=M_1-M_2\equiv\Delta M$.  Therefore, the sum over $M$ can be simplified as follows:

\begin{equation}
|{\cal M}_{\zeta}(\mathbf{k},\mathbf{0})|^2 = \frac{2\pi\hbar c}{Vk}|\langle J_1||T_L||J_2\rangle|^2  \langle J_2 M_2 L \Delta M|J_1M_1 \rangle^2  (-1)^{\Delta M-\zeta}\sum_{K=0}^{2L}\langle L-\Delta ML\Delta M|K0\rangle \langle L-\zeta L \zeta|K0\rangle   {\cal D}^{K}_{00}~.
\end{equation}

Lastly, by expressing $|\langle J_1||T_L||J_2\rangle|^2$ in terms of the mean life $\tau$ (Eq.~\ref{eqn:lifetime}) and noticing that ${\cal D}^{K}_{00}=P_K(\cos\beta)$, with $P_K$ being the Legendre polynomial of order $K$ ($\beta \in [0;\pi ]$), we have:

\begin{equation}
|{\cal M}_{\zeta}(\mathbf{k},\mathbf{0})|^2 = \frac{\pi\hbar^2 c}{2Vk^2\tau} (2L+1)\langle J_2 M_2 L \Delta M|J_1M_1\rangle^2  (-1)^{\Delta M-\zeta}
\sum_{K=0}^{2L}  \langle L-\Delta ML\Delta M|K0\rangle \langle L-\zeta L \zeta|K0\rangle P_{K}(\cos\beta)~,
\end{equation}

which is the expression used in the main text \cite{main}.

\section{Derivation of the operators time evolution equations}
In the following, we derive the time evolution equations, or equations of motion,  used for the calculations presented in the main text \cite{main}. The goal is to derive a set of three equations, describing the  time evolution of  the second quantization operators for photons, isomers $J^{\pi}=19/2^{-}$ and  the  nuclear  states $J^{\pi}=11/2^{+}$. We firstly derive the equations for these operators, ${c}_{\mathbf{k},\zeta}$, ${\Theta}_{\mathbf{q}}$ and ${X}_{\mathbf{q}}$ respectively; then we introduce $\tilde{c}_{\mathbf{k},\zeta}$, $\tilde{\Theta}_{\mathbf{q}}$, $\tilde{X}_{\mathbf{q}}$, thus ultimately leading to Eqs.~(6a-b) in \cite{main}.

Here, we calculate the time derivatives of the relevant operators, namely $\dot{c}_{\mathbf{k}}$, $\dot{\Theta}_{\mathbf{q}}$, $\dot{X}_{\mathbf{q}}$. To this purpose, it is convenient to introduce the Heisenberg representation. It is worth recalling that, in the Heisenberg picture, the evolution of an operator $\hat{L}$ is described by:

\begin{equation}
i\hbar \frac{\partial \hat{L}}{\partial t} = [\hat{L},H]~.
\end{equation}

Thus, in the case of the photonic operator, we have:

\begin{equation}
i \hbar \dot{c}_{ \mathbf{k},\zeta^\prime} = [c_{\mathbf{k},\zeta^\prime},H] = [c_{\mathbf{k},\zeta^\prime},H_\gamma]+[c_{\mathbf{k},\zeta^\prime},H_{A\gamma}]~, \label{eqn:cheisenberg}
\end{equation}

where $H=H_{A}+H_{\gamma}+H_{A\gamma}$ is the Hamiltonian of the system, comprising the contributions from the free Cs atoms ($H_{A}$, Eq.~(2) in \cite{main}), the photonic field ($H_{\gamma}$, Eq.~(3) in \cite{main}) and the interaction between the photonic field and the nuclear isomers ($H_{A\gamma}$, Eq.~(4) in \cite{main}).

 The first commutator in Eq.~\ref{eqn:cheisenberg} is:

\begin{equation}
[c_{\mathbf{k}^{\prime},\zeta^\prime},H_\gamma] = \sum_{{ {\bf k},}\zeta} \hbar\omega (\mathbf{k})\left[  c_{\mathbf{k}^{\prime},\zeta^\prime},c^{+}_{\mathbf{k},\zeta}c_{\mathbf{k},\zeta}\right] = \hbar \omega (\mathbf{k}^{\prime}) c_{\mathbf{k}^{\prime},\zeta^\prime}~.
\end{equation}

 The second commutator gives:

\begin{equation}
[c_{\mathbf{ k^\prime},\zeta^\prime},H_{A\gamma}] = \sum_{{ {\bf k},{\bf p},}\zeta} {\cal M}_{ \zeta} (\mathbf{k},\mathbf{p}) 
   \delta_{{\bf k} , {\bf k^\prime}} \delta_{\zeta,\zeta^\prime} \Theta_\mathbf{p} X^{+}_{\mathbf{p}- \hbar \mathbf{k}} =
\sum_{ {\bf p}}  {\cal M}_{\zeta^\prime} (\mathbf{ k^\prime},\mathbf{p}) \Theta_\mathbf{p} { X^{+}_{\mathbf{p}-\hbar \mathbf{k}^\prime}} ~.
\end{equation}

Thus, changing $k'$ back into $k$, one obtains:

\begin{equation}
 i{\hbar}\dot{c}_{\mathbf{ k},\zeta^\prime} ={\hbar}\omega (\mathbf{k}) c_{\mathbf{ k},\zeta^\prime}  +  \sum_{ {\bf p}}  {\cal M}_{\zeta^\prime} (\mathbf{k},\mathbf{p}) \Theta_\mathbf{p} { X^{+}_{\mathbf{p}- \hbar \mathbf{k}}}~.
\end{equation}

Now we move to $\dot{\Theta}_{\mathbf{q}}$:

\begin{equation}
i{\hbar}\dot{\Theta}_\mathbf{q} = \left[   \Theta_\mathbf{q}  , H  \right] = \left[   \Theta_\mathbf{q}  , H_A  \right] + \left[   \Theta_\mathbf{q}  , H_{A\gamma}  \right]~.
\end{equation}

By calculating the two commutators, we find:

\begin{equation}
\left[   \Theta_\mathbf{q}  , H_A  \right] =  \sum_{ {\bf p} } \left[  \Theta_\mathbf{q}      ,   \left(\frac{p^2}{2m}+{\hbar}\omega_0 \right)  \Theta^{+}_\mathbf{p}\Theta_\mathbf{p}      \right] = \left(\frac{q^2}{2m}+{\hbar}\omega_0 \right)  \Theta_\mathbf{q}~,
\end{equation}

and:

\begin{equation}
 \left[   \Theta_\mathbf{q}  , H_{A\gamma}  \right] = \sum_{ { {\bf k}}{ ,}{ {\bf p}}{ ,}    \zeta} {\cal M}^{*}_\zeta (\mathbf{k},\mathbf{p}) c_{\mathbf{k},\zeta} \left[\Theta_\mathbf{q},\Theta_\mathbf{p}^{+}\right] { X_{\mathbf{p}- \hbar \mathbf{k}}} =
                                                          \sum_{{ {\bf k}}{ ,}\zeta}   {\cal M}^{*}_\zeta (\mathbf{k},\mathbf{q})  c_{\mathbf{k},\zeta}   X_{\mathbf{q}-\hbar\mathbf{k}}~.
\end{equation}

Therefore, from the previous equations, we obtain the explicit form of the time derivative of the isomeric operator:

\begin{equation}
i\hbar\dot{\Theta}_\mathbf{q} =  \left(\frac{q^2}{2m}+{\hbar}\omega_0 \right)  \Theta_\mathbf{q} + \sum_{{ {\bf k},}\zeta} {\cal M}^{*}_\zeta (\mathbf{k},\mathbf{q})  c_{\mathbf{k},\zeta}   X_{\mathbf{q}-{ \hbar}\mathbf{k}}~.
\end{equation}

We now take into consideration the remaining $\dot{X}_{q}$. By following the same approach as for the other operators, we have:

\begin{equation}
i\hbar\dot{X}_\mathbf{q} = \left[   X_\mathbf{q}  , H  \right] = \left[   X_\mathbf{q}  , H_A  \right] + \left[   X_\mathbf{q}  , H_{A\gamma}  \right] ~. \label{eqn:xheisenberg}
\end{equation}

The explicit form of the two commutators is:

\begin{equation}
\left[   X_\mathbf{q}  , H_A  \right] = \sum_{{ {\bf p}}} \left[X_\mathbf{q} ,\frac{p^2}{2m}X^{+}_\mathbf{p}X_\mathbf{p}\right] = \frac{q^2}{2m}  X_\mathbf{q}
\end{equation}

\begin{equation}
\left[   X_\mathbf{q}  , H_{A\gamma}  \right] = \sum_{{ {\bf k},}\zeta}    {\cal M}_\zeta (\mathbf{k},\mathbf{q+k})  c^{+}_{\mathbf{k},\zeta}   \Theta_{\mathbf{q}+ \hbar\mathbf{k}}~.
\end{equation}

Now, by substituting these results into Eq.~\ref{eqn:xheisenberg}, we obtain:

\begin{equation}
 i\hbar\dot{X}_\mathbf{q} = \frac{q^2}{2m}  X_\mathbf{q}  + \sum_{{ {\bf k},}\zeta}    {\cal M}_\zeta (\mathbf{k},\mathbf{q+k})  c^{+}_{\mathbf{k},\zeta}   \Theta_{\mathbf{q}+\hbar\mathbf{k}}~.
\end{equation}

Finally, we introduce the new set of operators $\tilde{c}_{\mathbf{k},\zeta}$, $\tilde{\Theta}_{\mathbf{q}}$, $\tilde{X}_{\mathbf{q}}$:

\begin{subequations}
\begin{equation}
 \tilde{c}_{\mathbf{k},\zeta} = c_{\mathbf{\mathbf{k}},\zeta} \exp\left[ i\omega(\mathbf{k}) t\right]
\end{equation}

\begin{equation}
\tilde{\Theta}_\mathbf{q} = \Theta_\mathbf{q} \exp\left[ i \left(\dfrac{q^2}{2m \hbar}+\omega_0 \right) t\right]
\end{equation}

\begin{equation}
\tilde{X}_\mathbf{q} = X_\mathbf{q} \exp\left[ i  \dfrac{q^2}{2m \hbar}    t\right]
\end{equation}
\end{subequations}

Then, the equations of motion for the new operators are:

\begin{subequations}
\begin{eqnarray}
i \hbar \dot{\tilde{c}}_{\mathbf{k},\zeta} &=&  \sum_{{ {\bf q}}}   {\cal M}_{\zeta} (\mathbf{k},\mathbf{q}) \tilde{\Theta}_\mathbf{q} \tilde{X}^{+}_{\mathbf{q}-{ \hbar \mathbf{k}}}
 \exp\left[ { - i  \left( -\omega(\mathbf{k})+\frac{q^2}{2m \hbar}-\frac{ (\mathbf{q}-{ \hbar \mathbf{k}})^2}{2m \hbar}+\omega_0 \right)} t \right]  \label{eqn:32}\\
 i \hbar \dot{\tilde{\Theta}}_\mathbf{q}       &=&  \sum_{{{\bf k},}\zeta}    {\cal M}^{*}_\zeta (\mathbf{k},\mathbf{q})  \tilde{c}_{\mathbf{k},\zeta}   \tilde{X}_{\mathbf{q}-\hbar \mathbf{k}} \exp\left[ { -i \left(\omega(\mathbf{q}/\hbar)-\omega_0+\frac{(\mathbf{q}-\hbar\mathbf{k})^2}{2m\hbar}-\frac{q^2}{2m\hbar}\right) t}  \right]     \\
 i \hbar \dot{\tilde{X}}_\mathbf{q}       &=&  \sum_{{ {\bf k},}\zeta}    {\cal M}_\zeta (\mathbf{k},\mathbf{q}+\hbar \mathbf{k})  \tilde{c}^{+}_{\mathbf{k},\zeta}   \tilde{\Theta}_{\mathbf{q}+\hbar \mathbf{k}} \exp\left[-i\left( -\omega(\mathbf{q}/ \hbar)+\omega_0 - \frac{q^2}{2m\hbar}+\frac{(\mathbf{q}+{ \hbar \mathbf{k}})^2}{2m \hbar}\right)t\right]~, \nonumber \\ & & \label{eqn:34}
\end{eqnarray}
\end{subequations}

which are used in the main text to obtain, after simplification with the Bogoliubov  c-number approximation (Eq.~(5) in \cite{main}), the reduced time evolution equations (6a) and (6b).

\section{Solution of the reduced time evolution equations}
By using Eq.~(5) of \cite{main}, the evolution equations \ref{eqn:32}-\ref{eqn:34} reduce to:

\begin{subequations}
\begin{eqnarray}
 i \hbar \dot{\tilde{c}}_{\mathbf{k},\zeta}           &=&  \sqrt{n_0}  \xi_{\zeta} (\mathbf{k},\mathbf{0}) \tilde{X}^{+}_{{ -\hbar {\bf k}}} \exp[i \Delta_\mathbf{ k} t ] \label{eqn:ctildedot} \\
 i\hbar \dot{\tilde{X}}_{\hbar \mathbf{k}}       &=&  \sqrt{n_0}  \sum_{\zeta}  \xi_\zeta (\mathbf{-k},\mathbf{0})  \tilde{c}^{+}_{\mathbf{-k},\zeta}    \exp[i \Delta_\mathbf{- k} t ] 
\end{eqnarray} \label{eq:red_eq}
\end{subequations}

 where $n_0=N_0/V$ is the volume density of the isomeric BEC, $\xi_{\pm} (\mathbf{k},0)\equiv\sqrt{V}{\cal M}_{\pm}(\mathbf{k},0)$, and

\begin{equation}
\Delta_{\pm \mathbf{k}} =  \omega(\mathbf{\pm  k})+\dfrac{\hbar (\pm k)^2}{2m}-\omega_0~.\label{eqn:delta}
\end{equation}

The remaining explicit time-dependence in Eqs. \ref{eq:red_eq} can be eliminated by introducing the operator
\begin{equation}
\doubletilde{c}_{\mathbf{k},\zeta} = \tilde{c}_{\mathbf{k},\zeta} \exp\left[- i\Delta_{ {\mathbf k}}      t\right]~,
\end{equation}
 so that the relevant equations become
\begin{subequations}
\begin{eqnarray}
 i \hbar \dot{\doubletilde{c}}_{\mathbf{k},\zeta}           &=&  \hbar\Delta_\mathbf{k}\doubletilde{c}_{\mathbf{k},\zeta}+\sqrt{n_0}  \xi_{\zeta} (\mathbf{k},\mathbf{0}) \tilde{X}^{+}_{{ -\hbar {\bf k}}}  \label{eqn:ctildedot} \\
 i\hbar \dot{\tilde{X}}_{\hbar \mathbf{k}}       &=&  \sqrt{n_0}  \sum_{\zeta}  \xi_\zeta (\mathbf{-k},\mathbf{0})  \doubletilde{c}^{+}_{\mathbf{-k},\zeta}     
\end{eqnarray} \label{eq:red_eq_bis}
\end{subequations}

From Eqs.~\ref{eq:red_eq_bis}, then, we derive:

\begin{equation}
\hbar \ddot{\doubletilde{c}}_{\mathbf{k},\zeta}  + i \hbar \Delta_{\mathbf{k}}\dot{\doubletilde{c}}_{\mathbf{k},\zeta}- \frac{1}{\hbar } n_{0} \xi_{\zeta}(\mathbf{k}, \mathbf{0}) \sum_{\zeta^{\prime}} \xi^{*}_{\zeta^{\prime}}(\mathbf{{k}},\mathbf{0})\doubletilde{c}_{\mathbf{k},\zeta^\prime}  = 0~.
\end{equation}

Also from Eq. (\ref{eqn:ctildedot}) we can derive the initial condition for $\dot{\doubletilde{c}}_{\mathbf{k},\zeta} $:

\begin{equation}
\dot{\doubletilde{c}}_{\mathbf{k},\zeta}(0)   =   -i \Delta_\mathbf{k} \doubletilde{c}_{\mathbf{k},\zeta}(0)   -\frac{i}{\hbar}\sqrt{n_0 V} {\cal M}_{\zeta} (\mathbf{k},\mathbf{0}) \tilde{X}^{+}_{- { \hbar \mathbf{k}}} (0)~. 
\end{equation}

From the above equations, we then calculate the total number of  photons  $N_{\bf k }(t)$ emitted in the ${\bf k}$-mode as:

\begin{equation}
N_{\bf k} (t) = N_{{\bf k},-}+N_{{\bf k},+} =  \frac{2n_0}{\hbar^2 (\delta_0^2+\delta_1^2)}( |\xi_{-} (\mathbf{k},\mathbf{0})|^2 + |\xi_{+} (\mathbf{k},\mathbf{0})|^2)    \left[-\cos(\delta_0 t)+\cosh(\delta_1 t)\right]~,
\end{equation}

which is Eq.~(10) of the main text.

\section{Role of the intermediate state relaxation}

The inclusion of the fast relaxation of the 11/2$^{+}$ state leads to corrections in the coherent $\gamma$ emission characteristics. 
We assume that the the consequences of the rapid de-excitation of the intermediate state, at a rate $\Gamma_i$, can be evaluated by introducing a relaxation term of the type $\dot{\tilde{X}}_{\hbar \mathbf{k}} =  - \Gamma_i/2\tilde{X}_{\hbar\mathbf{k}}$ in the time-evolution equations, which read in this way

\begin{subequations}
\begin{eqnarray}
 i \hbar \dot{\doubletilde{c}}_{\mathbf{k},\zeta}           &=&  \hbar\Delta_\mathbf{k}\doubletilde{c}_{\mathbf{k},\zeta}+\sqrt{n_0}  \xi_{\zeta} (\mathbf{k},\mathbf{0}) \doubletilde{X}^{+}_{{ -\hbar {\bf k}}}  \label{eqn:ctildedot} \\
 i\hbar \dot{\tilde{X}}_{\hbar \mathbf{k}}       &=&  -i\hbar \frac{\Gamma_i}{2}\tilde{X}_{\hbar\mathbf{k}}+\sqrt{n_0}  \sum_{\zeta}  \xi_\zeta (\mathbf{-k},\mathbf{0})  \doubletilde{c}^{+}_{\mathbf{-k},\zeta}~.     
\end{eqnarray} \label{eq:red_eq_bis}
\end{subequations}

Proceeding as for the relaxation-free case, we find that the total photon number is given by

\begin{equation}
N_k(t) = \frac{
2 \exp(-\Gamma_i t/2)  n_0 ( |\xi_{-} (\mathbf{k},\mathbf{0})|^2 + |\xi_{+} (\mathbf{k},\mathbf{0})|^2)    (-\cos[\delta_0 t] + \cosh[\delta_1 t])
}{ (\delta_0^2+\delta_1^2) \hbar^2}
\end{equation}
with
\begin{equation}
 \sqrt{(2 \Delta + i \Gamma_i)^2 \hbar^2 - 
  16 n_0 \xi^2} = 
 2 \hbar \delta_0 + 2 i \hbar \delta_1 ~,\end{equation}
where we introduced the compact notation
\begin{equation}
\xi^2 \equiv  |\xi_{-}(\mathbf{k},\mathbf{0})|^2 + |\xi_{+} (\mathbf{k},\mathbf{0})|^2~.
\end{equation}

In the presence of relaxation of the intermediate state 11/2$^{+}$, the critical parameter is thus given by 
\begin{equation}
\tilde{\delta}_1 = \frac{1}{2}{\rm Im} \sqrt{(2 \Delta + i \Gamma_i)^2  - 
  16 \frac{n_0 \xi^2}{\hbar^2}} - \frac{1}{2}\Gamma_i~,
\end{equation}

and correspondingly the condition for exponential growth of the photon number is

\begin{equation}
 \frac{1}{2}{\rm Im} \sqrt{(2 \Delta + i \Gamma_i)^2  - 
  16 \frac{n_0 \xi^2}{\hbar^2}} - \frac{1}{2}\Gamma_i >0~.
\end{equation}

The expression for $\tilde{\delta}_1 $ can be rewritten as

\begin{equation}
\tilde{\delta}_1 =\frac{1}{2}{\rm Im} \left[
\sqrt{(2 \Delta + i \Gamma_i)^2  - 
  16 \frac{n_0 \xi^2}{\hbar^2}} - i\Gamma_i\right] 
\end{equation}

We notice that the  half-life of the intermediate state 11/2$^{+}$ has not been measured. The simplest assumption would be to use one Weisskopf unit, which gives a half-life 50 ps (for a 787 keV, E2 transition). Thus, for the specific case of large relaxation considered here, we can expand in series of of $16 \frac{n_0 \xi^2}{\hbar^2(\Delta+i \Gamma_i)^2}$. To the first order, we have:

\begin{eqnarray}
\tilde{\delta}_1 &=&\frac{1}{2}{\rm Im} 
\left[ (2 \Delta + i \Gamma_i)\sqrt{ 1 -  16 \frac{n_0 \xi^2}{\hbar^2(\Delta+i \Gamma_i)^2}} -i\Gamma_i \right]\\
&\simeq&\frac{1}{2}{\rm Im} 
\left[ (2 \Delta + i \Gamma_i)\left( 1 -  8 \frac{n_0 \xi^2}{\hbar^2(\Delta+i \Gamma_i)^2}\right) -i\Gamma_i \right]\\
&=&\frac{1}{2}{\rm Im} 
\left[ 2 \Delta  -  8 \frac{n_0 \xi^2}{\hbar^2 (\Delta+i \Gamma_i)} \right]\\
&=&
\frac{ 4 \frac{n_0 \xi^2}{\hbar^2} \Gamma_i  }
{(4\Delta ^2 + \Gamma^2_i )}
\end{eqnarray}

The relaxation of the lower state of the M4 $\gamma$ transition  thus leads to broadening of the gain region and reduction of gain peak.  In other words, the coherent photon emission is now spread over a range of frequencies determined by the J$^{\pi}=11/2^{+}$ state relaxation.  We can therefore conclude that the mechanism identified works also in the presence of relaxation, with the latter one determining the range of frequencies over which the emission is spread.

\section*{Acknowledegments}
This work was partially funded by the H2020-EU$.1.3.2$  programme through the Marie Curie Fellowship 2020-MSCA-IF-2014 ``GAMMALAS'' to L.~M.  (Proj. Ref. 657188), and by the Royal Society. P.~M.~W. acknowledges STFC support under grant no. ST/L005743/1.

\bibliography{isomers}